\documentclass[aps,twocolumn,amsmath,amssymb,preprintnumbers]{revtex4}

\usepackage{hyperref}
\usepackage{amsmath} \usepackage{amsfonts} \usepackage{amssymb}
\usepackage{graphicx}
\newcommand{\be}{\begin{equation}}
\newcommand{\ee}{\end{equation}}
\newcommand{\ba}{\begin{eqnarray}}
\newcommand{\ea}{\end{eqnarray}}
\newcommand{\nn}{\nonumber}
\newcommand{\beq}{\begin{equation}}
\newcommand{\eeq}{\end{equation}}
\newcommand{\beqa}{\begin{eqnarray}}
\newcommand{\eeqa}{\end{eqnarray}}

\newcommand{\bea}{\begin{eqnarray}}
\newcommand{\eea}{\end{eqnarray}}

\newcommand {\apgt} {\ {\raise-.5ex\hbox{$\buildrel>\over\sim$}}\ }
\newcommand {\aplt} {\ {\raise-.5ex\hbox{$\buildrel<\over\sim$}}\ }

\def\s0#1#2{\mbox{\small{$ \frac{#1}{#2} $}}}
\def\0#1#2{\frac{#1}{#2}}

\newcommand{\metric}{$g_{\mu\nu}$}

\begin{document}

\title[ ]{Dilaton Quantum Gravity}

\author{T. Henz} 
\affiliation{Institut f\"ur Theoretische
  Physik, Universit\"at Heidelberg, Philosophenweg 16, 69120
  Heidelberg, Germany} 

\author{J. M. Pawlowski} 
\affiliation{Institut f\"ur Theoretische
  Physik, Universit\"at Heidelberg, Philosophenweg 16, 69120
  Heidelberg, Germany}

\author{A. Rodigast} 
\affiliation{Institut f\"ur Theoretische
  Physik, Universit\"at Heidelberg, Philosophenweg 16, 69120
  Heidelberg, Germany}

\author{C. Wetterich} 
\affiliation{Institut f\"ur Theoretische
  Physik, Universit\"at Heidelberg, Philosophenweg 16, 69120
  Heidelberg, Germany} 
\begin{abstract}
We propose a simple fixed-point scenario in the renormalization flow of a scalar dilaton coupled to gravity. This would render gravity non-perturbatively renormalizable and thus constitute a viable theory of quantum gravity. On the fixed point dilatation symmetry is exact and the quantum effective action takes a very simple form. Realistic gravity with a nonzero Planck mass is obtained through a nonzero expectation value for the scalar field, constituting a spontaneous scale symmetry breaking. Furthermore, relevant couplings for the flow away from the fixed point can be associated with a ``dilatation anomaly'' that is responsible for dynamical dark energy. For the proposed fixed point and flow away from it the cosmological ``constant'' vanishes for asymptotic time. 
\end{abstract}

\maketitle

During recent years, much evidence has been collected for gravity to be asymptotically safe \cite{weinberg1979ultraviolet}, allowing for a non-perturbative quantization \cite{Reuter1}. Functional renormalization group methods \cite{Wetterich92,Reuter1994181} have been used to show that the ultraviolet fixed point has remained rather robust for many extended truncations beyond Einstein-Hilbert gravity \cite{Reuter:2012id, Percacci:2007sz, ReuterRV,  Litim:2011cp,percacci1}, thereby supporting and extending the results from various different investigations of the asymptotic safety scenario \cite{Hamber:2009mt, Ambjorn:2009ts}. Most recently, the attention was drawn to establishing a smooth trajectory from the ultraviolet regime towards the infrared limit \cite{Donkin:2012ud,Christiansen:2012rx, Nagy:2012rn}. The main ingredient of the fixed point is the ultraviolet scaling of the Planck mass proportional to the renormalization scale $k$.

For an exact fixed point dilatation, or scale, symmetry is exact. In general this entails an appropriate rescaling of the fields, usually including anomalous dimensions. The scale symmetry should be reflected in the quantum effective action in the limit where the infrared cutoff $k$ is sent to zero. To date, the form of the quantum effective action associated with the fixed point in Einstein-Hilbert gravity is still disputed. Furthermore, in this theory the flow of a relevant parameter away from the fixed point is needed in order to end up with a non-vanishing Planck mass. 

In this note we consider the system of a scalar field coupled to gravity. In the presence of a dilaton field $\chi$ it is straightforward to construct effective actions with dilatation symmetry \cite{Fujii:1982ms,Wetterich:1987fm,Shaposhnikov:2008xi}. We will argue that a corresponding rather simple fixed point may be present in the functional flow equations for the dilaton coupled to gravity. Such a fixed point would define ``dilaton quantum gravity''. The proposed fixed point is associated to functions of $y=\chi^2/k^2$ that do not change during the flow. In this sense it involves infinitely many couplings of the dilaton-gravity system. 

We investigate the functional flow within a simple truncation for the euclidian effective average action,
\be\label{Gamma4d}
\Gamma_k =\int d^4x \sqrt{g} \left( V_k(\chi^2)-\frac12 F_k(\chi^2)\,R+\frac{1}{2}g^{\mu\nu}
\partial_\mu\chi\partial_\nu\chi \right),
\ee
where $R$ is the curvature scalar. The potential $V_k(\chi^2)$ and the coupling function $\frac 1 2 F_k(\chi^2)$ are generic analytic functions of the square of the scalar field $\chi$. The renormalization group flow of an action of the type \eqref{Gamma4d} was first investigated in \cite{percacci1}. As a first approach, we use a simple kinetic term for $\chi$, leaving a more complex form for future work. At a later stage, other extensions, such as higher derivative terms as $R^2$, may be included. The effective action $\Gamma_k$ is not scale invariant, in contrast to the setting of Ref. \cite{Codello:2012sn}. In case of a fixed point, however, scale invariance is obtained as a result of fluctuations in the limit $k\to 0$.

Our aim is a study of the flow of the functions $V_k$ and $F_k$ as a function of the infrared cutoff $\sim k$. Effectively, this cutoff ensures that only quantum fluctuations with (covariant) momenta $|q|\gtrsim k$ are included in the computation of the effective average action $\Gamma_k$. In four dimensions, $\chi$  has mass-dimension one and we introduce dimensionless functions $v_k(y)$ and $f_k(y)$ through 
\be\label{split} 
V_k=k^4\,y^2\, v_k(y),\,\, F_k=k^2\,y\, f_k (y),\,\,\text{where}\,\, y=\frac{\chi^2}{k^2}.
\ee
Our main suggestion is that the flow of the truncated effective action \eqref{Gamma4d} has a fixed point for which the dimensionless functions $v$ and $f$ become independent of the renormalization scale $k$. They obey the asymptotic properties \be\label{dimlessasymptotics}
\lim_{y\to\infty}f(y)=\xi\quad\text{and} \quad\lim_{y\to\infty}v(y)=0.
\ee

Such a fixed-point scenario has striking consequences. At the end, we are interested in the limit $k\to 0$, where all quantum fluctuations are included.  For any nonzero value of $\chi$, this corresponds to $y\to\infty$, such that the quantum effective action for the dilaton-gravity system takes the simple form 
\be\label{GammaFP}
\Gamma=\int d^4x \sqrt{g} \left(\frac{1}{2}g^{\mu\nu}\partial_\mu\chi\,\partial_\nu\chi- \frac 1 2 \xi \chi^2 \,R \right).
\ee
This action is dilatation symmetric, as it should be for an exact fixed point. For any nonzero expectation value of $\chi$ dilatation symmetry is spontaneously broken, thereby generating a physical mass scale. The associated Goldstone boson is massless and will be referred to as the dilaton. This scenario is realized by cosmological solutions of the field equations derived from \eqref{GammaFP} as outlined in \cite{Wetterich:1987fm}. The action \eqref{GammaFP} describes a viable theory of gravity (see Ref. \cite{Wetterich:2013jsa} for a recent discussion). In an extended setting with matter, particle masses are also proportional to $\chi$, such that all experimental constraints are obeyed.

Indeed, performing a canonical Weyl scaling of the metric \metric\ and using a rescaled scalar field $\phi$, the effective action \eqref{GammaFP} becomes the Einstein-Hilbert action coupled to a massless scalar field, namely 
\be
\Gamma=\int d^4x \sqrt{g}  \left(\frac{1}{2}g^{\mu\nu}\partial_\mu\phi\,\partial_\nu\phi- \frac 1 2 M^2 \,R \right).
\ee
The scale $M$ characterizes the spontaneous breaking of dilatation symmetry, $\langle \chi\rangle =M/\sqrt{\xi}$. In this normalization $M$ equals the reduced Planck mass, related to Newtons constant by $G^{-1}_N=8\pi M^2$. Thus the physical content of the effective action \eqref{GammaFP} is Einstein gravity coupled to a massless dilaton. The dilaton plays no role in late cosmology, unless further ``dilatation anomalies'' are generated which can play the role of dynamical Dark Energy \cite{Wetterich:1987fm}, see below.

We are interested in the behavior of possible fixed-point solutions for large $y$. In this region the properties of the fixed point are rather simple. The main ingredient will be the simple observation that for large $y$ and nonzero 
\be
f_0=\lim_{y\to\infty}f_k(y)
\ee
the strength of the gravitational interaction is given by $f_0^{-1}\chi^{-2}$. For the gravity induced flow of the dimensionless quantities only the dimensionless combination $f_0^{-1}y^{-1}$ can be of relevance. However, this quantity vanishes for $y\to\infty$ and the gravitational interactions are absent in this limit. For $y\to\infty$ and $v\,(y\to\infty)\to 0$ one is then left with a free scalar field. In turn, for a free scalar field the flow cannot induce a nontrivial $\chi$-dependent effective potential, such that only a constant term can flow in $V_k$. For large $y$, the leading term in $v_k$ is then proportional to $y^{-2}$ and vanishes for $y\to\infty$, such that at the fixed point 
\be\label{7}
\lim_{y\to\infty}v(y)=v_{-2}\,y^{-2}.
\ee
For a vanishing strength of the gravitational interaction the leading term in the gravitational sector of the effective action does not flow either. Thus $f_0$ does not depend on $k$, establishing the asymptotic behavior \eqref{dimlessasymptotics} with $\xi=\lim_{k\to 0}f_0$. Considering the fixed point \eqref{dimlessasymptotics} we will find that the dominant correction to $F_k$ is a term $\sim k^2$, resulting in an asymptotic form for large $y$ given by 
\be\label{asymf} 
\lim_{y\to\infty} f(y)=\xi+f_{-1}\,y^{-1}.
\ee

We emphasize that these simple properties hold only for $y\to\infty$. For small $y$ the effective coupling of the scalar to the graviton induces interactions as shown explicitly in Ref.  \cite{Eichhorn:2012va}.

\medskip\noindent
{\em Flow equations.} In order to quantify the results of the previous section, let us now turn to the explicit flow equations for the dimensionless functions $f_k(y)$ and $v_k(y)$. The flow at constant values of $y$ takes the general form $\big(t=\ln(k/\mu)\big)$
\begin{eqnarray}
\nonumber
\partial_t v_k(y) &=&\,2\,y\, v_k^\prime(y)  +\frac{1}{y^2}\,\zeta_V, \label{fullflowVa} \\[0.5cm]
 \partial_t f_k(y) &=&\,2\,y\, f_k^\prime(y) + \frac{1}{y}\,\zeta_F. \label{fullflowFa}
\end{eqnarray}
For the computation of $\zeta_V$ and $\zeta_F$ we employ the infrared cutoff introduced in Ref. \cite{percacci1}, which is based on Ref. \cite{litimopt1}. This also applies to other technical aspects such as a standard background field treatment, which is the same as in Ref. \cite{percacci1}, and we use deDonder gauge. For the truncation \eqref{Gamma4d} we find
\begin{eqnarray}\label{zetaV}
\zeta_V&=& \frac{1}{192\pi^2}
\Biggl\{
6 + \frac{30 \, \tilde V}{\Sigma_0} + \frac{3 (2   \Sigma_0 + 24 \, y \,  \tilde F^{\prime} \, \Sigma_0^{\prime} +  \, \tilde F \Sigma_1) }{\Delta}\nn\\
&&+\delta_V\Biggr\},\nn
\end{eqnarray}
\begin{eqnarray}
\zeta_F&=&
\frac{1}{1152\pi^2}
\Biggl\{ 
150 + \frac{30  \, \tilde F \, ( 3 \, \tilde F - 2 \tilde V )}{\Sigma_0 ^2} \\
&& - \frac{12}{\Delta} \left( 24 \, y  \, \tilde F^{\prime} \, \Sigma_0^{\prime} +  2\Sigma_0 +   \tilde F \Sigma_1 \right)-6 
y \, (3 \, \tilde F^{\prime 2} + 2\Sigma_0^{\prime 2} )
\nn\\\notag
&&- \frac{36}{\Delta ^2} 
\Biggl[
2 y \, \Sigma_0 \, \Sigma_0^{\prime} \, ( 7 \, \tilde F^{\prime} - 2\tilde V^{\prime} ) \, 
( \Sigma _1 - 1)+ 2 \, \Sigma_0^2 \, \Sigma_2\\\notag
\end{eqnarray}
\begin{eqnarray}
&&+ 2 \, y \Sigma _1 \, ( 7  \, \tilde F^{\prime} - 2\tilde V^{\prime} ) \, ( 2 \, \Sigma_0 \, \tilde V^{\prime} - \tilde V \, \Sigma_0^{\prime} )\nn\\
&&+ 24 \, y \, \tilde F^{\prime}  \, \Sigma_0 \, \Sigma_0^{\prime} \, \Sigma_2 - 12\, y \, \tilde F \,  \Sigma_0^{\prime 2} \, \Sigma_2
\Biggr] +\delta_F\Biggr\}.\nn \label{eq12}
\end{eqnarray}
Here we employ
\begin{eqnarray}
\tilde V=y^2\,v_k(y)~,~\tilde F&=&y\, f_k(y),\nn\\
\Sigma_0 = \frac12 \tilde F-\tilde V ~,~  \Delta &=& \left( 12 \, y  \, 
\Sigma_0^{\prime2} + \Sigma_0 \, \Sigma_1 \right),\nn\\[6pt]
\Sigma_1 = 1+ 2 \, \tilde V^{\prime} + 4 \, y \, \tilde V^{\prime \prime}~,~
\Sigma_2 &=& \tilde F^{\prime} + 2 \, y \, \tilde F^{\prime \prime}.
\end{eqnarray}
The contributions
\ba\label{zetaF}
\delta_V&=&
\left( \frac{4}{\tilde F} + \frac{5}{2\Sigma_0} + 
\frac{ \Sigma_1}{2\Delta} \right)  \left(\partial_t \tilde F +2\,\tilde F  - 2\, y \,\tilde F^{\prime}\right) \nn\\
&&+ \frac{ 12 \, y  \, \Sigma_0^{\prime}}{ \Delta} \,\left( \partial_t \tilde F^{\prime} -2\,y\,\tilde F^{\prime\prime}\right),\nn\\
\delta_F&=&- \frac{\partial_t \tilde F + 2\,\tilde F - 2\, y \,\tilde F^{\prime}}{\tilde F}
\Biggl[
30  -  \frac{5 \tilde F \, (7 \, \Sigma_0 + 4 \, \tilde V ) }{\Sigma_0^2} \nn\\\notag
&&+ \frac{3}{\Delta ^2}
\Biggl(
  \tilde F \, \Sigma_1 \, \Delta 
+ 8 \, y \, \tilde V^{\prime} \, \Sigma_0^{\prime} \, \Delta   -24   \, y  \, \tilde F  \, \Sigma_0^{\prime 2} \, \Sigma_2\nn\\
&&-2y  \, \tilde F \, \Sigma_0^{\prime} \, \Sigma_1 (7 \, \tilde F^{\prime}-2\tilde V^{\prime})\Biggr)\Biggr]\\
&&+  \, \frac{6  \, y }{\Delta ^2} \Biggl[
(   \, \tilde F^{\prime} + 10 \, \tilde V^{\prime} ) \Delta  - 24  \, \Sigma_0 \, \Sigma_0^{\prime} \, \Sigma_2 \nn\\
&&- 2 \, ( 7   \tilde F^{\prime} - 2\tilde V^{\prime} ) \, \Sigma_0 \, \Sigma_1 
\Biggr]\,\left( \partial_t \tilde F^{\prime} -2\,y\,\tilde F^{\prime\prime}\right),\nn
\ea
arise from the field dependence in the cutoff. They vanish for $y\to\infty$. Neglecting these contributions altogether does not change the structure of the results obtained. The flow Eqs. \eqref{fullflowFa} are in complete accordance to what was found in \cite{percacci1}.

The flow generators $\zeta_V$ and $\zeta_F$ are expressed in terms of $v_k(y)$ and $f_k(y)$ as well as their first and second derivatives with respect to $y$, denoted by primes. They also involve $y$ explicitly. The terms $2\,y\, v_k^\prime(y)$ and $2\,y\, f_k^\prime(y)$ arise in Eqs. \eqref{fullflowFa} when transforming from flow equations at fixed $\chi$ to the corresponding equations at fixed $y$, with $\partial_t y\bigm|_\chi=-2\, y$,
\begin{equation}\label{AA}
\frac{1}{\chi^2}\partial_t F_k\bigm|_\chi=\partial_t f_k\bigm|_\chi=\partial_t f_{k}\bigm|_y+f'_k\partial_ty\bigm|_\chi,
\end{equation}
and similarly for $\partial_t v_k$. The functions $\zeta_V$ and $\zeta_F$ are defined as
\begin{equation}\label{AB}
\zeta_V=\frac{1}{k^4}\partial_t V_k\bigm|_\chi~,~\zeta_F=\frac{1}{k^2}\partial_tF_k\bigm|_\chi.
\end{equation}

We first investigate the generators $\zeta_V$ and $\zeta_F$ in the limit $y\to\infty$. In this limit we expand $v$ and $f$ in inverse powers of $y$
\begin{eqnarray}\label{D2a}
v(y)&=&v_0+v_{-1}y^{-1}+v_{-2}y^{-2}+\dots,\nonumber\\
f(y)&=&f_0+f_{-1}y^{-1}+\dots.
\end{eqnarray}
Eqs. \eqref{zetaV} yield
\begin{eqnarray}\label{D2b}
\lim_{y\to\infty}\zeta_V=\bar\zeta_V~,~
\lim_{y\to\infty}\zeta_F=\bar\zeta_F,
\end{eqnarray}
where the limits depend on $f_0$ and $v_0$. We concentrate on 
\begin{eqnarray}\label{D2d}
v_0&=&v_{-1}=0,\quad\text{where}\nonumber\\
\bar\zeta_V&=&\frac{3}{32 \pi ^2}+\frac{5+33 f_0}{96 \pi ^2  (1+6 f_0)}\frac{\partial_tf_0}{f_0},\nonumber\\
\bar\zeta_F&=&\frac{77+534 f_0}{192 \pi ^2 (1+6 f_0)}\nonumber\\
&&+\frac{17+186 f_0+720 f_0^2 }{576 \pi^2 (1 +6  f_0)^2}\frac{\partial_tf_0}{f_0}.
\end{eqnarray}

We are interested in fixed-point solutions $f^*(y),v^*(y)$ for which $\partial_t f_k(y)=\partial_t v_k(y)=0$. The contributions $\sim \partial_t f_0$ vanish in this case. In the limit $y\to\infty$ Eqs. \eqref{fullflowFa} are then easily solved by
\begin{eqnarray}\label{AE}
\lim_{y\to\infty}f^*(y)=\xi+\frac{\bar\zeta_F}{2y}~,~
\lim_{y\to\infty}v^*(y)=\frac{\bar\zeta_V}{4y^2}.
\end{eqnarray}
This coincides with the expectations \eqref{7}, \eqref{asymf}, with $v^*_{-1}=0$ and $v_{-2}^*=\bar \zeta_V/4$, $f_{-1}^*=\bar \zeta_F/2$.

\medskip\noindent
{\em Scaling Solution.} 
The most obvious global fixed-point solution to the flow Eqs. \eqref{fullflowFa} is given by setting both $\tilde V$ and $\tilde F$ equal to a constant. This corresponds to Einstein-Hilbert gravity with an additional scalar field $\chi$, which couples to gravity only through the metric \metric. This solution obeys Eq. \eqref{fullflowVa} exactly for all $y$ and is numerically given by
\be
\tilde V = y^2 v(y) = 0.008620 \quad \text{and} \quad \tilde F = y f(y) = 0.04751.
\ee

Starting in the close vicinity of this fixed point we find that the latter is not stable as $k$ is lowered. The behavior of small deviations from the fixed point
 is governed by a linear differential equation that can be cast into the form of eigenvalue equations for the growth rate. For the $y$-dependence of its solutions we find an exponential growth for large $y\gg 1$ while for small $y\ll 1$ the typical behavior is $\propto \sqrt{y}$.

\begin{figure}
 \includegraphics[width=\columnwidth]{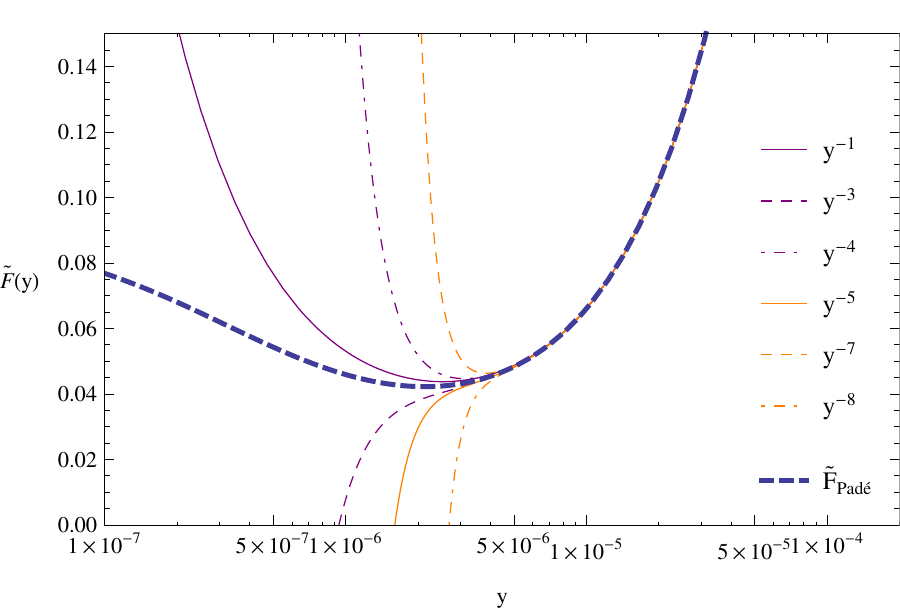}
\caption{Taylor expansions for $\tilde F(y)$ at $y=\infty$ with $\xi=4000$, truncated at $y^0$ to $y^{-8}$ and the Pad\'e improvement including powers of $y^{-4}$ in both numerator and denominator. The splitting up of the Taylor series at the radius of convergence points to a breakdown of perturbation theory.\label{fig:taylorlarge}}
\end{figure}

In order to gain insight into other classes of solutions, we have performed an expansion of $f^*(y)$ and $v^*(y)$ in powers of $y^{-1}$ including the order $y^{-8}$. The result depends on the value $\xi$ which is not fixed at this stage. The series shows excellent apparent convergence for $y\geq y_0 = 1/(10|\xi|)$. The convergence for smaller $y$ can be improved by a Pad\'e approximation, which we call $\tilde V_\text{Pad\'e}$ and $\tilde F_\text{Pad\'e}$, respectively. We show the result in figure \ref{fig:taylorlarge}, where numerator and denominator are expanded up to order $y^{-4}$. The Pad\'e approximation satisfies the fixed-point equations to good accuracy with a squared relative error of $<0.1$ for $y>y_0\approx 1/(50 \xi)$.

\begin{figure}
 \includegraphics[width=\columnwidth]{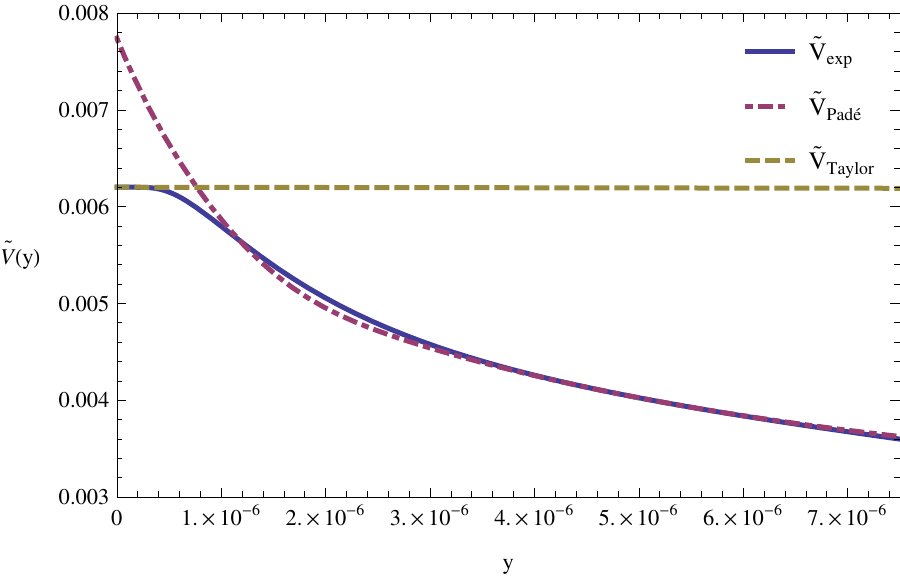}\\
 \includegraphics[width=\columnwidth]{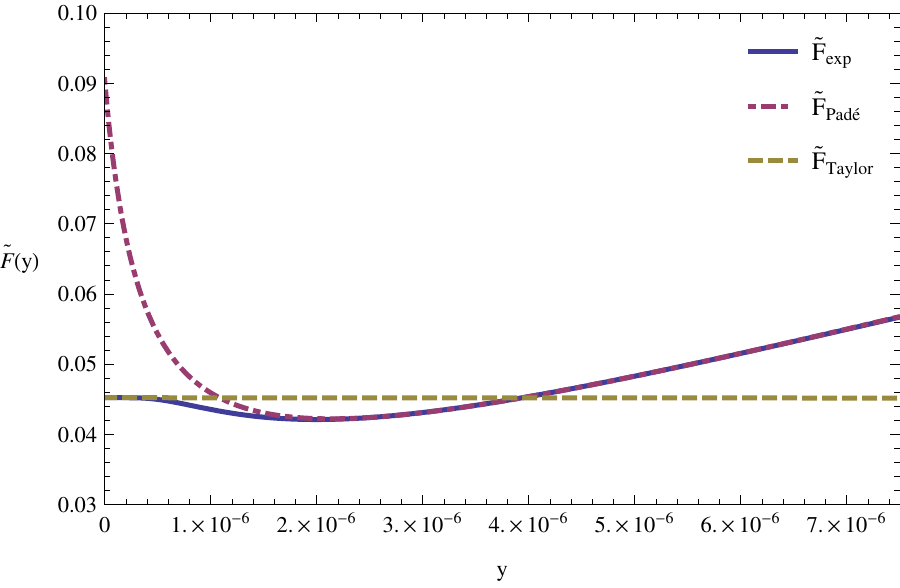}
\caption{Taylor expansion at $y=0$ with $V_0=6.2 \times 10^{-3}$ and $F_0=2.3 \times 10^{-2}$, the Pad\'e approximation at $y=\infty$ with $\xi=4000$ and the exponential ansatz \eqref{eq:exp_ansatz} matched at $y=5 \times 10^{-6}$ for the fixed functions $\tilde V(y)$ and $\tilde F(y)$. \label{fig:fixed_fcts}}
\end{figure}

We emphasize that the flow Eqs. \eqref{fullflowFa} are reliable only if the relevant inverse propagators $\Sigma_0$ and $\Delta$ in the spin $2$ and spin $0$ sector remain positive for all $y$. These quantities correspond to the inverse graviton and scalar propagators in the presence of the cutoff $k$. For $y\to\infty$ one has $\Sigma_0=\frac12\xi y,\Sigma_1=1$, $\Delta=\frac12\xi(1+6\xi)y$ such that $\Sigma_0$ and $\Delta$ are positive provided $\xi>0$. We have checked that the positivity of $\Sigma_0$ and $\Delta$ also holds for the Pad\'e approximation for the values of $\xi$ shown in figure \ref{fig:taylorlarge} and \ref{fig:fixed_fcts}.

The positivity requirement for the propagators singles out asymptotic solutions for $y\to\infty$ for which $v_0\leq 0$. In fact, the asymptotic fixed-point solutions of Eqs. \eqref{fullflowFa} also have solutions with $v_0\neq 0$, with 
\begin{eqnarray}\label{D2c}
&\bar\zeta_V=&-\frac{1}{48\pi^2}\left(6-\frac{\partial_t f_0}{f_0}\right),\nonumber\\
&\bar\zeta_F=&\frac{1}{1728\pi^2}\left(249-41\frac{\partial_t f_0}{f_0}\right),
\end{eqnarray}
and corresponding values for the coefficients in the $y^{-1}$-expansion $v_{-2}^*=-0.00317$ and $f_{-1}^*=0.0073$. (One also finds constant values for $\bar\zeta_V$ and $\bar\zeta_F$ in case of $f_0=0$. They differ from Eqs. \eqref{D2c} and \eqref{D2d}.) For $v_0\neq 0$ the asymptotic form $\tilde V=v_0y^2$, $\tilde F=\xi y$ implies a negative $\Sigma_0$ if $v_0>0$, rendering the propagation of the graviton unstable. With $\Sigma_0=-v_0 y^2,\Sigma_1=12 v_0y$, one has $\Delta=36 v^2_0y^3$ which remains positive for arbitrary $y\neq 0$.

Furthermore, we require that the potential $V_k$ in our ansatz \eqref{Gamma4d} is bounded from below in order to describe a stable theory. This holds for an asymptotic behavior with $v_0\geq 0$. Combining the two requirements of a positive inverse propagator and a bounded potential only the asymptotic behavior $v_0=0$ is left. This absence of a term $\sim v_0y^2$ is the crucial ingredient for the absence of a cosmological constant after Weyl scaling in Eq. \eqref{GammaFP}. 

The region of small $y\lesssim y_0$ is more difficult to access. One may investigate a Taylor expansion around $y=0$ for the functions $\tilde F=yf^*(y)$ and $\tilde V=y^2v^*(y)$,
\ba\label{X1}
\tilde F&=&yf=F_0+F_1y+F_2y^2+\dots\nn\\
\tilde V&=& y^2v=V_0+V_1y+V_2y^2+\dots
\ea
The fixed-point equations
\ba\label{X2}
\zeta_F-2\tilde F+2y\tilde F'&=&0,\nn\\
\zeta_V -4\tilde V+2 y\tilde V'&=&0,
\ea
require that the constant terms in $\zeta_F$ and $\zeta_V$ equal $2F_0$ and $4V_0$, respectively. In turn, these constants $\zeta^{(0)}_F$ and $\zeta^{(0)}_V$ involve $F_0,F_1$ and $V_0,V_1$. The system is not closed, and this property extends to higher orders in the Taylor expansion. For given $F_0$ and $V_0$ the Taylor expansion shows apparent convergence and we have expanded up to $y^5$. We matched the Taylor expansion to the Pad\'e expansion by the ansatz
\begin{equation}\label{eq:exp_ansatz}
\begin{aligned}
 \tilde F_{\text{exp}}(y)&= \tilde{F}_{\text{Taylor}}(y) + e^{-c/y}\tilde{F}_e(y),\\
  \tilde V_{\text{exp}}(y)&= \tilde{V}_{\text{Taylor}}(y) + e^{-c/y}\tilde{V}_e(y),
\end{aligned}
\end{equation}
where $\tilde{F}_{\text{Taylor}}$, $\tilde{V}_{\text{Taylor}}$ are the Taylor expansions to order $y^5$ and $\tilde{F}_e$, $\tilde{V}_e$ are polynomial functions of $y$ to order $y^2$. The coefficients of $\tilde{F}_e$ and $\tilde{V}_e$ are determined by matching the values and derivatives of $F$ and $V$ at some fixed $y=y_0$. We show the result in figure \ref{fig:fixed_fcts}, for which we have determined $F_0$ and $V_0$ as well as the parameter $c$ by varying their values around the limits of the corresponding Pad\'e approximations for $y\to 0$ and optimizing the result as an approximate solution to Eq. \eqref{fullflowVa}.

The solution shown in figure \ref{fig:fixed_fcts} is only approximate. Further analysis will be needed to establish a global scaling solution. It is well conceivable that an exact fixed point exists only for one particular value of $\xi$. At the present stage we cannot exclude that such a fixed-point solution is connected smoothly to the Einstein-Hilbert fixed point as $y$ goes to zero. The true fixed point could also occur for a negative sign of the scalar kinetic term, or for a function $K(y)$ multiplying this term which changes sign as a function of $y$. For the time being we assume that an exact fixed point can finally be established and we explore further consequences of this scenario. 

\medskip\noindent
{\em Flow Away from the Fixed Point.} In contrast to the asymptotic freedom scenario in Einstein gravity there is no need to depart from the fixed point for the realization of a realistic model of gravity. In Einstein gravity the Planck mass is an intrinsic scale that violates dilatation symmetry. It is generated by the flow departing from the fixed point. In the language of critical phenomena the squared Planck mass corresponds to a relevant parameter. For a trajectory exactly on the fixed point the Planck mass would vanish. In contrast, in dilaton gravity a trajectory exactly on the fixed point can well describe gravity. Dilatation symmetry is then exact and this version of quantum gravity would correspond to a regularization that preserves exact dilatation symmetry \cite{Wetterich:1987fm,Bezrukov:2012hx}. The Planck mass is generated by spontaneous dilatation symmetry breaking through the expectation value $\langle\chi\rangle$.

Nevertheless, the exact realization of the fixed-point trajectory is not necessary. We may also discuss models of gravity where the renormalization flow starts very close to the fixed point in the ultraviolet, but the trajectory ultimately departs from the exact fixed point. This is analogous to asymptotic freedom in non-abelian gauge theories where the gauge coupling goes to zero for an infinite momentum scale (Gaussian fixed point), while for any finite arbitrarily large value of the scale it has a nonzero value. We will next show that this type of trajectory also leads to realistic gravity. It even entails an interesting cosmology with dark energy of the quintessence type.

We are interested in the range of large $y$ or small $k^2\ll \chi^2$. Deviations from the fixed-point flow are most easily investigated by flow equations at fixed $\chi$. A good approximation to the flow equations for $V$ and $F$ is then given by Eq. \eqref{AB} with constant $\bar\zeta_V$ and $\bar\zeta_F$, namely
\be\label{5B}
\partial_t V = \bar\zeta_V k^4~,~\partial_t F = \bar\zeta_F k^2.
\ee
We look for solutions of Eqs. \eqref{5B} which respect the asymptotically constant values of $\zeta_V$ and $\zeta_F$. Solutions of this type are simply
\begin{eqnarray}
V &=& \frac{\bar\zeta_V}{4} k^4 + \bar V, \nonumber\\\label{5C}
F &=& \xi\chi^2 + \frac{\bar\zeta_F}{2} k^2 + \bar F.
\end{eqnarray}
Indeed, the $\chi^2$-independent integration constants $\bar V$ and $\bar F$ do not influence the value of $\bar\zeta_V$ and $\bar\zeta_F$. These constants contribute to $\zeta_V$ and $\zeta_F$ only in the order $y^{-1}$.  This contrasts with the addition of a term $\bar m^2 \chi^2$ to the potential $V$. Even though this would formally still constitute a solution to Eq. \eqref{5B}, the values of $\bar \zeta_V$ and $\bar \zeta_F$ would be altered for the range $\bar m^2/k^2>\xi$, since the asymptotic behavior of $\Sigma_0$ and $\Delta$ for $y\to\infty$ is modified. We observe that $\bar V$ is a relevant parameter. The value of the dimensionless ratio $V/k^4$ is dominated by $\bar V$ for $k\to 0$. In the limit $k\to 0$ the integration constant $\bar V$ appears as a type of cosmological constant in the Jordan frame. Also $\bar F$ remains important for $k\to 0$, but this is limited to $\xi\chi^2\lesssim \bar F$. Both $\bar V$ and $\bar F$ introduce explicit mass scales and break dilatation symmetry. For $\
\bar F$ and $\bar V$ different from zero the range of small $\chi$ with $\xi\chi^2\lesssim \bar F$ resembles strongly the setting of Einstein-Hilbert gravity. The interesting new aspects of the present work concern the range $\xi\chi^2\gg \bar F$.

In the presence of nonzero $\bar V$ and $\bar F$ the effective action of dilaton quantum gravity becomes for $k\to 0$
\be\label{5D}
\Gamma=\int d^4 x \sqrt{g} \left(\frac{1}{2}g^{\mu\nu}\partial_\mu\chi\,\partial_\nu\chi- \frac 1 2 (\xi \chi^2+\bar F) \,R +\bar V \right),
\ee
generalizing Eq. \eqref{GammaFP}. A cosmological constant in the Jordan frame has cosmological consequences that differ strongly from a cosmological constant in the Einstein frame \cite{Wetterich:1987fm,Wetterich:2013jsa}. This can be seen by a Weyl scaling 
\be
g'_{\mu\nu} = \frac{\xi\chi^2 + \bar F}{M^2}\,g_{\mu\nu},\nonumber 
\ee 
combined with a rescaling of the scalar field 
\be
\varphi=M\ln\left(\frac{\xi\chi^2+\bar F}{M^2}\right).
\ee
In terms of the new variables the effective action \eqref{5D} reads 
\ba
\Gamma &=& \int d^4 x\sqrt{g}
\left\{-\frac{M^2}{2}R+\frac12 k^2(\varphi)\partial^\mu\varphi\partial_\mu\varphi+V(\varphi)\right\},\nn\\
k^2(\varphi)&=&\frac{1}{4\xi}\left[1+6\xi+\left(\frac{M^2}{\bar F}\exp\left(\frac\varphi M\right)-1\right)^{-1}\right].
\ea
The cosmon potential 
\be\label{cp}
V(\varphi)=\bar V\exp \left(-\frac{2\varphi}{M}\right)
\ee
decreases exponentially for large $\varphi/M$. Adding radiation and matter the cosmology described by the associated field equations admits a typical scaling solution for large $\varphi/M$ with a fixed fraction of dynamical dark energy or quintessence. This holds for a negative sign of the scalar kinetic term in Eq. \eqref{Gamma4d}, with a suitable range of $F$ for which the model remains stable. Our discussion can be extended to this case. For solutions where $\varphi$ goes to infinity for asymptotic time  the ``cosmological constant'' vanishes asymptotically. 

We conclude that a fixed point with the asymptotic behavior \eqref{dimlessasymptotics} shows highly interesting properties. It would result in non-perturbatively renormalizable quantum gravity with a scale invariant quantum effective action \eqref{GammaFP}. This action contains no term $\sim\chi^4$ due to $v_0=0$. The absence of such a term ensures the vanishing of the cosmological constant in the Einstein frame. Furthermore, deviations from exact scale invariance can lead to dark energy cosmology with an asymptotically vanishing ``cosmological constant''. These interesting aspects may motivate the substantial work that is required in order to firmly establish such a fixed point.

\vfill

\bibliographystyle{bibstyle}

\end{document}